\documentclass[%
reprint,
superscriptaddress,
amsmath,amssymb,
aps,
prb,
]{revtex4-1}

\usepackage{sidecap}
\sidecaptionvpos{figure}{c}

\usepackage{graphicx}% Include figure files
\usepackage{dcolumn}% Align table columns on decimal point
\usepackage{bm}% bold math
\usepackage[
colorlinks=true,
linkcolor=blue,
citecolor=blue,
filecolor=blue,
urlcolor=blue]{hyperref}

\usepackage{amsmath,amsfonts,
	amsgen,amsbsy,amsopn,amstext,amssymb
}
\usepackage{epstopdf}
\usepackage{color}
\renewcommand{\i}{i}

\begin{document}
	
	\preprint{}

\title{Dynamic Nonlinear Focal Shift in Amplitude Modulated Moderately Focused Acoustic Beams}

\author{No\'e Jim\'enez}
\affiliation{Laboratoire d'Acoustique de l'Universit\'e du Maine - UMR CNRS 6613, Av. O. Messiaen, 72085 Le Mans, France}
\author{Francisco Camarena}
\affiliation{Instituto de Instrumentaci\'on para Imagen Molecular (I3M), Centro Mixto UPV CSIC CIEMAT, Camino de Vera s/n, 46022, Valencia, Spain}
\author{Nuria Gonz\'alez-Salido}
\affiliation{ITEFI, Spanish National Research Council (CSIC), Serrano 144, 28006 Madrid, Spain}
\date{\today}

\begin{abstract}
	The phenomenon of the displacement of the position of the pressure, intensity and acoustic radiation force maxima along the axis of focused acoustic beams under increasing driving amplitudes (nonlinear focal shift) is studied for the case of a moderately focused beam excited with continuous and 25 kHz amplitude modulated signals, both in water and tissue. We prove that in amplitude modulated beams the linear and nonlinear propagation effects coexist in a semi-period of modulation, giving place to a complex dynamic behaviour, where the singular points of the beam (peak pressure, rarefaction, intensity and acoustic radiation force) locate at different points on axis as a function of time. These entire phenomena are explained in terms of harmonic generation and absorption during the propagation in a lossy nonlinear medium both, for a continuous and an amplitude modulated beam. One of the possible applications of the acoustic radiation force displacement is the generation of shear waves at different locations by using a focused mono-element transducer excited with an amplitude modulated signal.
\end{abstract}
\maketitle

\section{Introduction}

The study of the acoustic field generated by focusing sources in nonlinear regime is a continuously developing field of research as finite amplitude sound beams are increasingly used in medicine and industry\cite{Duck1998,Blitz1995,Lucas1982,Bakhvalov1978,Duck1986}. Nonlinear propagation implies asymmetric wave steepening, progressive harmonic generation, nonlinear absorption, sound saturation, self-refraction and self-demodulation \cite{Hamilton1998}. All these nonlinear effects change the spatial distribution of the acoustic field respect to the linear case, i.e., among other things, the location of the on-axis maximum and minimum pressure, intensity and acoustic radiation force (ARF), as well as the focal spot dimensions. 

The nonlinear focal shift phenomenon, defined as the shift of the maximum pressure (and also intensity and ARF) position along the axis of focused acoustic beams under increasing driving voltages, has been discussed and interpreted in previous works. In 1980 Bakhvalov et al.\cite{Bakhvalov1980} predicted a shift in the position of the on-axis pressure maximum in unfocused beams where a migration of the location of the maximum was shown, first away from, and then towards the transducer, as the exciting voltage of the source was increased. Duck and Starritt\cite{Duck1986} (1986) studied this phenomenon in slightly focused sources as those used in commercial medical pulse-echo equipments, showing that the nonlinear focal shift exists for on-axis maximum and minimum pressure, with different behaviour. Averkiou and Hamilton\cite{Averkiou1997} (1997) observed this phenomenon experimentally in a moderately focused piston (linear gain $G=p/p_0 = 10.36$ with $p$ is the value of the pressure at the geometrical focus and $p_0$ the pressure at the surface of the transducer). The nonlinear focal shift phenomenon was reported by Makov et al.\cite{Makov2008} in low gain transducers, and discussed it in terms of the harmonics nonlinearly generated during the propagation of a finite amplitude wave. They provided also experimental evidence of the nonlinear shift in slightly focused transducers ($G=4$). Bessonova et al.\cite{Bessonova2009} reported a numerical study where the nonlinear focal shift is shown for a moderately focused piston ($G=10$) in a range of intensity covering both the shift of the maximum pressure towards the geometrical focus at first, even passing beyond the focus, and then the shift backwards to the transducer. They also provided an interpretation of the phenomenon based on the self-defocusing effect due to the asymmetrical distortion of the wave profile and to the increase in propagation velocity of the compressive phase of the wave close to the beam axis. Recently, Camarena et al.\cite{Camarena2013} proved experimentally that at high amplitudes and for moderately focusing ($G=18.8$) the position of the on-axis pressure maximum and radiation force maximum can surpass the geometrical focal length.

The location of the singular points of a focused ultrasonic beam, i.e., the on-axis maximum and minimum pressure, maximum intensity and ARF, depends on the nonlinear degree of the propagated waves. This is especially relevant in moderately focused beams since the focusing is high enough to make the nonlinear effects relevant, but at the same time the transversal area of the focus is not as small as in highly focused devices, making the self-refraction effect to play an important role\cite{Hamilton1998,Camarena2013}. The singular points of a beam (as for example the location of the maximum ARF) generated by a moderately focused mono-element emitter can be moved just by growing the amplitude of the voltage applied to the source. One of the applications that come to mind is to generate supersonic shear waves with a focused mono-element transducer.

The aim of this work is to investigate experimentally and numerically the dynamic behavior of the singular points of a moderately focused beam operating from linear to nonlinear regime, and excited with amplitude modulated (AM) and continuous signals. Both, water and soft tissue (human liver) media has been considered. The paper is structured as follows: in Sec. \ref{Sec2} the experimental set-up and methods are described, providing a linear characterization of the beam and adjusting the source parameters to obtain the numerical results. Section \ref{Sec3} shows the experimental and numerical results for the maximum focal displacement in water (Sec. \ref{Sec31}) and the dynamical focal shifts for a 25 kHz AM beam (Sec. \ref{Sec32}). In Section Sec. \ref{Sec33} the study is numerically extended to propagation in soft tissue (human liver). Finally, concluding remarks are given in Sec. \ref{Sec4}.

\section{Materials and methods}\label{Sec2}
\subsection{Experimental set-up}\label{Sec21}
\begin{figure}[tbp]
	\centering
	\includegraphics[width=8cm]{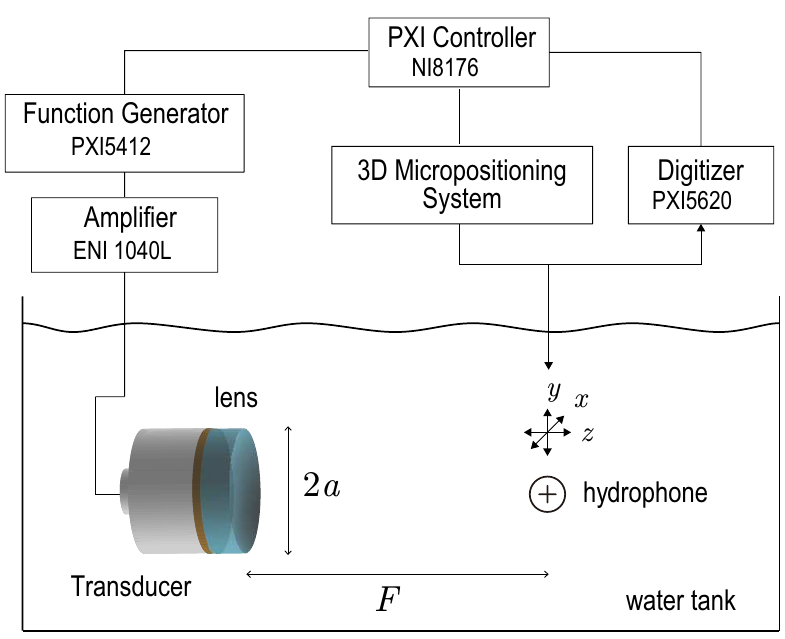}
	\caption{Scheme of the experimental set-up for the pressure measurement in water.}
	\label{fig:1}
\end{figure}

The experimental set-up for the pressure measurements in water follows the classical scheme of confronted emitting focused source and receiving calibrated membrane hydrophone, located inside a $0.75\times0.6\times0.5$ m water tank filled with degassed and distilled water at $26^\circ$, as shown in Fig.~\ref{fig:1}. The ultrasound source was formed by a plane single element piezoceramic crystal (PZ 26, Ferroperm Piezoceramics, Denmark) mounted in a custom designed stainless-steel housing and a poly-methyl methacrylate (PMMA) focusing lens with aperture $2a=50$ mm and radius of curvature $R=70$ mm. The resonant frequency of the transducer was $f_0=1.112$ MHz, and it was driven by a signal generator (14 bits, 100 MS/s, model PXI5412, National Instruments) and amplified by a linear RF amplifier (ENI 1040L, 400 W, +55 dB, ENI, Rochester, NY). The pressure field was measured by a PVDF membrane hydrophone with a 200 $\mu$m active diameter (149.6 mV/MPa sensitivity at 1.112 MHz, Model MHB-200, NTR/Onda) calibrated from 1MHz to 20 MHz). The hydrophone signals were digitized at a sampling rate of 64 MHz by a digitizer (model PXI5620, National Instruments) averaged 500 times to increase the signal to noise ratio. An $x$-$y$-$z$ micro-positioning system (OWIS GmbH) was used to move the hydrophone in three orthogonal directions with an accuracy of 10 $\mu$m. All the signal generation and acquisition process was based on a National Instruments PXI-Technology controller NI8176, which also controls the micro-positioning system. Temperature measurements were done over the whole process ensuring no temperature changes of $\pm0.6^\circ$ C.

\subsection{Linear characterization of the beam}\label{Sec22}
To determinate accurately the position of the radiator axis, a variant of the procedure described in Cathignol et al.\cite{Cathignol1997} was developed. Firstly, the transducer was oriented along the $z$-axis of the micro-positioning system. In order to find the focal region of the transducer, the maximum pressure distribution generated by a 20-cycles sinusoidal pulsed burst ($\mathrm{V}_0 = 6$ Vpp) was measured along the axis of the radiator. These measurements provided a rough estimation of the transducer focal area. Then, the pressure waveforms $p(t,x,y,z)$ were measured at the focal area in five planes along the $z$ axis of the micro-positioning system, separated $\Delta z=5$ mm. These planes were transversal to the $z$ axis, $6\times6$ mm ($x$-$y$ planes) and waveforms were acquired with 0.5 mm spatial resolution on them (144 measurement points per plane). here, a zero-gain band-pass filter was applied to each waveform. The maximum for each waveform was selected by adjusting a Gaussian function to the histogram of maxima in the tone burst. The equipressure curves in each plane built with the selected maxima typically had a circular form: This was indicative of good axial symmetry of the radiator. Finally, the coordinates ($x$-$y$) of the maximum and the z position of each transversal plane were used for fitting a 3D line that determines the acoustic axis. 

\begin{figure}[tbp]
	\centering
	\includegraphics[width=8cm]{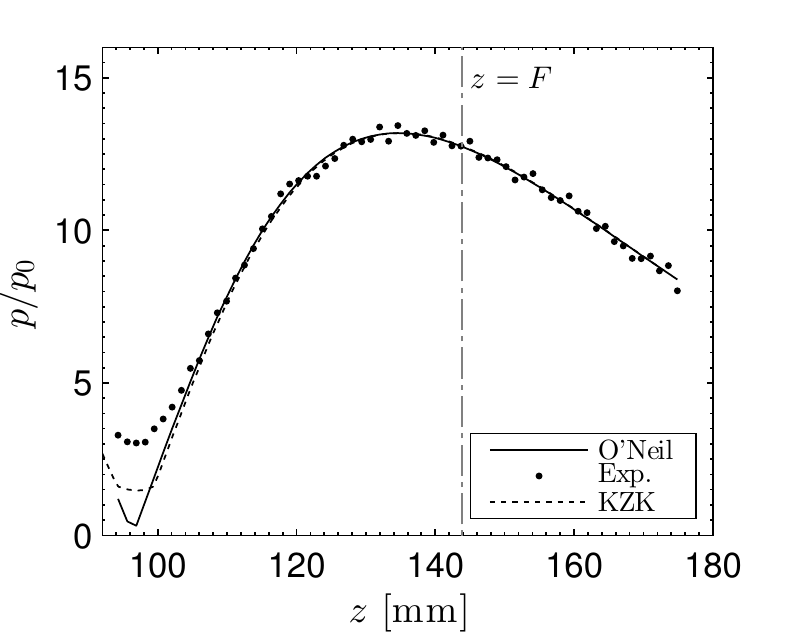}
	\caption{On-axis maximum pressure distribution for small amplitude wave propagation obtained by experimental data (dots), O'Neill solution (solid line) and KZK numerical solution (dotted line). The vertical dashed-dotted line shows the position of the geometric focus.}
	\label{fig:2}
\end{figure}

A set of 63 signals was measured over the previously determined radiator axis (dots in Fig. \ref{fig:2}; 6 Vpp initial voltage). Zero-phase band-pass filtering was employed here in order to reduce noise and do not introduce any temporal delay in the signals. Then, time of flight ($t_f$) was measured at each point over the axis to locate the absolute hydrophone position ($z_0'$) with respect to the radiator. It was fitted as $z' = c_0t_f + z_0'$ , where $z'$  was the axial position of the different points, $c_0$  the estimated sound speed in pure water. The least squares error for this fit ensures accuracy better than $\pm0.7$ mm for absolute focal estimations.

The geometrical focal length (considering the PMMA focusing lens) and the aperture of the transducer were nominally stated by the manufacturer as $F=157$ mm and $2a=50$ mm, respectively. The fit of the analytic O'Neil solution\cite{ONeil1949} to the experimental data is shown in Fig.~\ref{fig:2} and provides an effective aperture of the transducer of 55 mm and an effective geometrical focal length of $F=143.9$ mm. Finally, the KZK simulation\cite{Lee1993} provides an effective aperture of the transducer of 56 mm and a geometrical focal length of 143.9 mm. 

The results of both models, the O'Neil and the calculated with the ``best fit'' aperture and geometrical focal length in the KZK simulation, are in good agreement with the experimental data, as shown in Fig.~\ref{fig:2}. The peak pressure location in linear regime is located at $z = 135.6$ mm, which corresponds to a linear focal shift of $\Delta F = -8.6$ mm respect the geometric focus. This initial shift is used to obtain the focal shifts in nonlinear regime.

\subsection{Measurement procedure}\label{Sec23}
To study the nonlinear focal shift produced in continuous and AM beams, the acoustic field on the radiator axis was measured for different initial pressures. Firstly, the transducer was excited by a 40 cycles-sine wave burst with voltages in the range from 5 $\mathrm{V}_p$ to 228 $\mathrm{V}_p$ (a total of 27 values). Then, the experiment was repeated for an AM beam generated by the 25 kHz modulation of a continuous beam (90 cycles-sine wave burst). For all the input voltages, the axial field generated was measured at 63 points over the radiator axis with a spatial resolution of 1.3 mm in order to evaluate the position of the on-axis pressure maximum with accuracy sensitive to the nonlinear pressure focal shift phenomenon (estimated in less than 1 cm from numeric simulations\cite{Makov2008}). 

The on-axis intensity distribution, $I(z)$, has been evaluated by using the temporal pressure waveforms $p(t,z)$ as:
\begin{equation}\label{eq:I}
I(z) = \frac{1}{nT}\int_{t_0}^{t_0+nT}\frac{p^2(z,t)}{\rho_0 c_0} \,,
\end{equation}
\noindent where $T$ is the period, $n$ an integer, $\rho_0$ the ambient density and $c_0$ the sound speed.

As the transversal dimension of the beam is much smaller than the characteristic absorption length, the transversal component of the ARF in the focal area is practically negligible \cite{Bakhvalov1987,Pishchalnikov2002}. Therefore the axial component of the ARF in a viscous heat-conducting medium can be evaluated from:
\begin{equation}\label{eq:ARF}
\mathrm{ARF}_z(z) = \frac{\delta}{\rho_0^2 c_0^5}\left\langle \left( \frac{\partial p(t,z)}{\partial t} \right)^2\right\rangle\,,
\end{equation}

\noindent where $\delta$ is the sound diffusivity and the angular brackets denote temporal averaging over fast acoustic oscillations. 

On the other hand, the axial ARF in a soft-tissue, modeled as a frequency-power law attenuation medium, has been calculated as
\begin{equation}\label{eq:ARFI}
\mathrm{ARF}(z) = \frac{2}{c_0}\int_{\omega=0}^{\infty}\alpha_0\omega^\gamma I(\omega,z)\,,
\end{equation}

\noindent where the frequency dependent attenuation is given by $\alpha(\omega)=\alpha_0\omega^\gamma$, with $\alpha_0$ and $\gamma$ the attenuation coefficient and exponent of the frequency power law respectively, and $I(\omega,z)$ the intensity of frequency component. This relation must be corrected by a term $c_0^{-2}\partial I/\partial t$ accounting for the temporal modulation of the signal \cite{guzina2015}, but in this study due to the different order of magnitude of the carrier and modulation frequencies, this term can be neglected. Note the validity of this relation is restricted at the focus of moderately focused beams where the waves can be considered quasi-planes.

Finally, to obtain the experimental focal displacements, the axial distribution of maximum/minimum pressure, intensity and ARF curves was fitted to a polynomial of degree four over the focal area. The estimated position of the maxima provides the absolute focal displacements, were the errors committed were below $\pm1$ mm for all the present cases.

\subsection{Numerical Model}\label{Sec24}
Numerical modelling of the experimental conditions was performed by using the numerical solution of the KZK nonlinear parabolic equation over a cylindrical axisymmetric coordinate system $\mathbf{r} (r,z)$. This model takes into account the nonlinearity, diffraction (assuming a beam in the paraxial/parabolic approximation), thermo-viscous absorption and relaxation \cite{Hamilton1998}.

For propagation along the positive $z$-axis, the dimensionless KZK equation in a thermoviscous and relaxing media is written as \cite{Lee1993,yang2005}
\begin{equation}\label{eq:kzk}
\begin{split}
\frac{\partial P}{\partial \sigma} = & \frac{1}{4G}\int_{-\infty}^{\tau} \left( \frac{\partial^2 P}{\partial\rho^2} + \frac{1}{\rho}\frac{\partial P}{\partial \rho} \right)d\tau + A\frac{\partial^2 P}{\partial \tau^2} + N P\frac{\partial P}{\partial \tau} \\ &+\sum_n D_n \int_{-\infty}^{\tau}\frac{\partial^2 P}{\partial \tau'^2}e^{-(\tau-\tau')/\theta_n}d\tau'\,, 
\end{split}
\end{equation}
\noindent where $P=p / p_0$ is the pressure normalized to the pressure at the source plane $p_0$, $\sigma=z/F$ the dimensionless axial coordinate with $F$ the source geometric focal, $\rho=r/a$ the dimensionless radial coordinate with $a$ the source radius, $\tau=\omega_0 t'$ the dimensionless retarded time with $\omega_0$ the beam angular frequency, $t' = t - z / c_0$ the retarded time, $G = z_d / F$ the gain with $z_d = \omega_0a^2/2c_0$ the characteristic diffraction distance (i.e. the Rayleigh distance), $A=F/z_a$ the absorption parameter with $z_a=1/\alpha$ the characteristic absorption distance and $\alpha=\delta\omega_0^2/2/c_0^3$ the thermoviscous absorption coefficient, $N=F/z_s$ the parameter of nonlinearity with $z_s = \rho_0c_0^3/\beta p_0 \omega_0$ the shock formation distance for a plane-wave, and $\beta$ the coefficient of nonlinearity of the medium. The relaxation parameters are $\theta_n=\omega_0 t_n$ and $D_n=k_0Fc_n'/c_0$, where $t_n$ and $c_n'$ are the characteristic relaxation time and the small-signal sound speed increment for the $n$-th relaxation process, respectively.

For simulations in water, i.e., thermo-viscous fluids, Eq.~(\ref{eq:kzk}) was solved using the time-domain algorithm developed by Lee \cite{Lee1993}, the so called KZK-Texas code, and relaxation was neglected. For simulations in liver, i.e., frequency power-law attenuation media, Eq.~(\ref{eq:kzk}) was solved numerically using the KZK-Texas code including multiple relaxation and heterogeneities in an axisymmetric domain \cite{yang2005}. The relaxation parameters were optimized to fit the tissue frequency dependent attenuation and dispersion \cite{yang2005}, i.e., a frequency power-law attenuation and its corresponding dispersion. Three relaxation processes were used. Both numerical approaches were based on operator splitting, where each term was solved sequentially for each $\Delta\sigma$ step. Thus, the diffraction term, i.e., first right hand side term in Eq.~(\ref{eq:kzk}), was solved using an implicit backward finite difference (IBFD) method for $\sigma<0.1$ and a Crank–Nicolson finite difference (CNFD) method thereafter. Same methods were used for the thermo-viscous absorption and relaxation terms, i.e., second and fourth RHS terms in Eq.~(\ref{eq:kzk}). Finally, the nonlinearity, i.e., third RHS term in Eq.~(\ref{eq:kzk}) was solved analytically using the Poisson solution where it was checked that the nonlinear distortion will not allow the wave form to become multivalued. The $\Delta\sigma$ step was decreased accordingly to avoid multivalued waveforms. Finally, heterogeneities were included accordingly to Ref. \cite{yang2005} Eq.(15) by a plane wave pressure transmission
coefficient and a temporal phase shift in the retarded frame, accounting for the propagation through layers with different density and sound speed. 

Simulations were performed in water and tissue, both with continuous and amplitude modulated beams. Seventy values of $p_0$ were used in each case, ranging from 5 kPa to 105 kPa. The lower initial pressure was obtained from the previously linear characterization of the beam (see Sec. \ref{Sec22}). Simulation parameters were $c_0=1488$ m/s, $\rho_0=998$ kg/m3, $\beta=3.5$, $\delta=5.13 \cdot 10^{-6}$  m$^2$/s, $\alpha=0.19$ dB/m, $F=143.9$ mm and $a=28$ mm. This parameters lead to the dimensionless factors $G=12.75$, $A=3.87 \cdot 10^{-3}$, and the parameter of nonlinearity ranging between $5.24 \cdot 10^{-3}< N < 0.125$. Therefore, the Gol'dberg ratio, $\Gamma=z_a/z_s=N/A$, ranges between $0.9 < \Gamma < 21.3$, so nonlinearity dominates over absorption effects. On the other hand, the Khokhlov number, $K=z_s/z_d=1/NG$, ranges between $15 > K > 0.63$, so the nonlinearity dominates over diffraction effects only for high amplitudes. In the case of simulations including tissue, a layer of liver tissue was introduced at $z=70$ mm. The parameters were $c_\mathrm{liv}=1600$, $\rho_\mathrm{liv}=1060$ and $\beta=4.35$; where the parameters of 3 relaxation processes were optimized to model a frequency power law attenuation $\alpha=\alpha_0\omega^\gamma$ with $\alpha=0.5$ dB/cm/MHz$^\gamma$ and $\gamma=1.1$. Both, the retrieved attenuation and its corresponding dispersion using relaxation agrees with the frequency power-law from .5 to 40 MHz.

The source condition was modeled by applying a temporal delay across the source to account for the focusing as:
\begin{equation}
\left.P\left(\tau,\rho\right)\right|_{\sigma=0} = p_s\left(\tau + G \rho^2\right) g\left(\rho\right) \,, 
\end{equation}
\noindent where $p_s$ is the temporal waveform and for an uniform piston, $g(\rho)$ is the step function defined as $g(\rho)=1$ for $0\le \rho \le 1$, and $g(\rho)=0$ for $\rho>1$.

For the continuous acoustic beam, the temporal waveform was expressed as
\begin{equation}
p_s(t) = p_0 \mathrm{e}^{-\left( \frac{\omega_0 t}{M\pi}\right)^m}\sin(\omega_0 t) \,.
\end{equation}
\noindent where the power $m$ was chosen to simulate a nearly rectangular pulse with continuous amplitude, and the denominator $M$ indicates the number of cycles simulated. Here $m=8$ and $M=25$ were used.

In the case of AM beams, the temporal waveform used was
\begin{equation}
p_s(t) = p_0 \mathrm{e}^{-\left( \frac{\omega_0 t}{\omega_0/\omega_m\pi}\right)^m}\sin(\omega_0 t)\sin\left( \omega_m t \right)  \,.
\end{equation}
\noindent Here, we simulated only four AM packets, i.e. only 2 cycles of the modulation frequency, $f_m = \omega_m/2\pi = 25$ kHz.

\section{Results}\label{Sec3}
\subsection{Maximum nonlinear focal shift in water}\label{Sec31}

\begin{figure}[tbp]
	\centering
	\includegraphics[width=8cm]{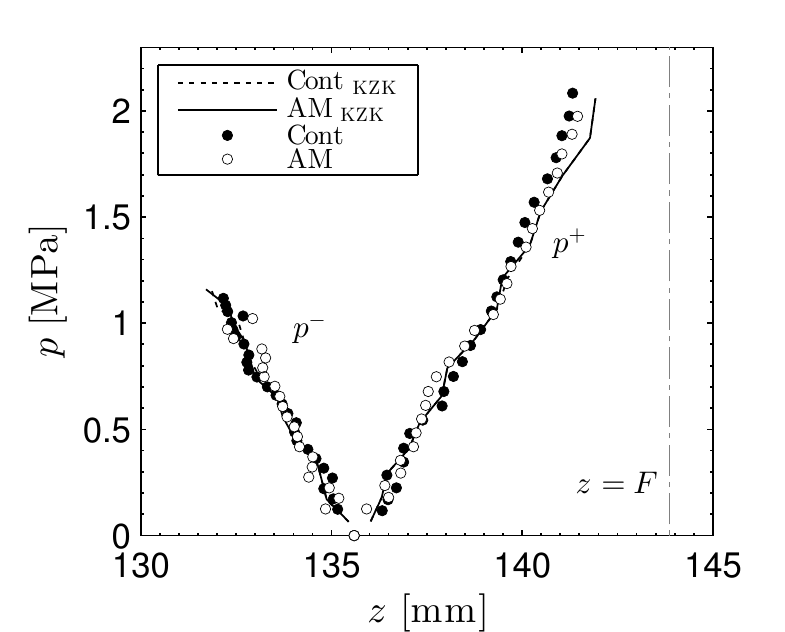}
	\caption{On axis peak pressure location for both, continuous and 25 kHz-AM beams. Experimental and KZK solutions are showed. Location of the pressure maximum/minimum of experimental pulsed beam (black circles), experimental AM beam (open circles), simulated pulsed (continuous line) and AM beam (dashed line). The dashed-dotted vertical line represents the position of the geometric focus, $F=143.9$ mm.}
	\label{fig:3}
\end{figure}

Figure \ref{fig:3} shows the value and location of the on-axis maximum and minimum pressures for 27 spaced initial excitation voltages at the source (between 5 $\mathrm{V}_p$ to 228 $\mathrm{V}_p$). These particular points exhibit similar nonlinear shifts in the experiment and simulation for both continuous and 25 kHz-AM beams. The on-axis peak pressure position moves away from the transducer $+$5.8 mm for experimental data and 6.4 mm for simulated data. On the other hand, the on-axis minimum pressure position moves towards the transducer a maximum displacement of $-$4.0 mm in both cases. This focal displacements lead to a total difference between peak pressure and rarefaction focal positions of 9.8 mm for experimental measurements and 10.4 mm for simulation results, which agree with the results reported in Camarena et al.\cite{Camarena2013} and Makov et al.\cite{Makov2008} for a focused beam with Fresnel number 4.06. The agreement between experiments and simulations in the quasi-linear region (low and medium input voltages) is good, but they differ slightly at high power levels (nonlinear regime), being the maximum pressure nonlinear focal shift effect slightly higher in the simulation. A similar discrepancy was also detected in Camarena et al.\cite{Camarena2013} and explained according to several reasons: first, the frequency response of the hydrophone is bounded to 20 MHz, which limits the number of harmonics detected by the hydrophone. Second, the sound field does not present a flat and uniform distribution over the active area of the receptor (200 $\mu$m active diameter), thus the measure is underestimated after the spatial averaging of the measurement region, which does not happen in the simulation. Finally, another possible source of error is due to the non-uniform vibration of the transmitter, as discussed before. These hypotheses have been discussed in detail by Canney et al.\cite{Canney2008}. 

\begin{figure}[tbp]
	\centering
	\includegraphics[width=8cm]{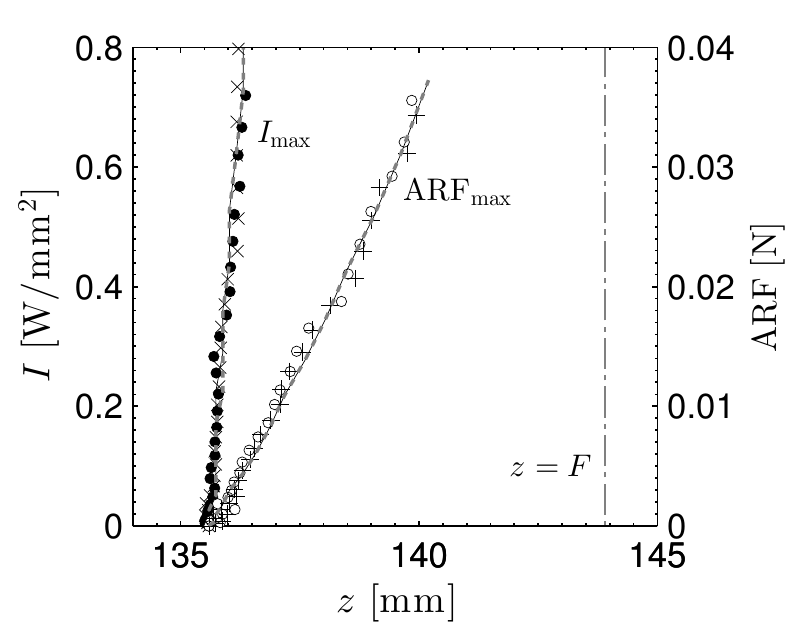}
	\caption{On axis maximum intensity position for continuous ($\times$) and AM beams (black circles). Location of the on axis ARF peak for continuous ($+$) and AM beams (white circles). Simulation curves are superimposed for continuous (continuous line) and AM beams (dashed lines). The dashed-dotted vertical line represents the position of the geometrical focus.}
	\label{fig:4}
\end{figure}

In the case of the maximum intensity location (see Fig.~\ref{fig:4}), again the nonlinear shift is equal in both experiments, with the continuous and the 25 kHz-AM beams. The intensity, evaluated from Eq.~(\ref{eq:I}) reaches a maximum at 80 W/cm$^2$ and suffers a shift of $+1$ mm both in experiment and KZK simulation, which is very small compared with the $+5.8$ mm displacement of the maximum pressure. Nevertheless the ARF distribution (evaluated from Eq.~(\ref{eq:ARF})) undergoes a nonlinear shift that follows the same trend than pressure (compare Fig.~\ref{fig:3} with Fig.~\ref{fig:4}), although the displacements of the maximum ARF are always slightly lower than the measured for pressure. A $+4.6$ mm ARF shift compared to $+5.8$ mm maximum pressure shift is reached in the highest nonlinear regime, i.e. 2.1 MPa peak pressure.

\begin{figure}[tbp]
	\centering
	\includegraphics[width=8cm]{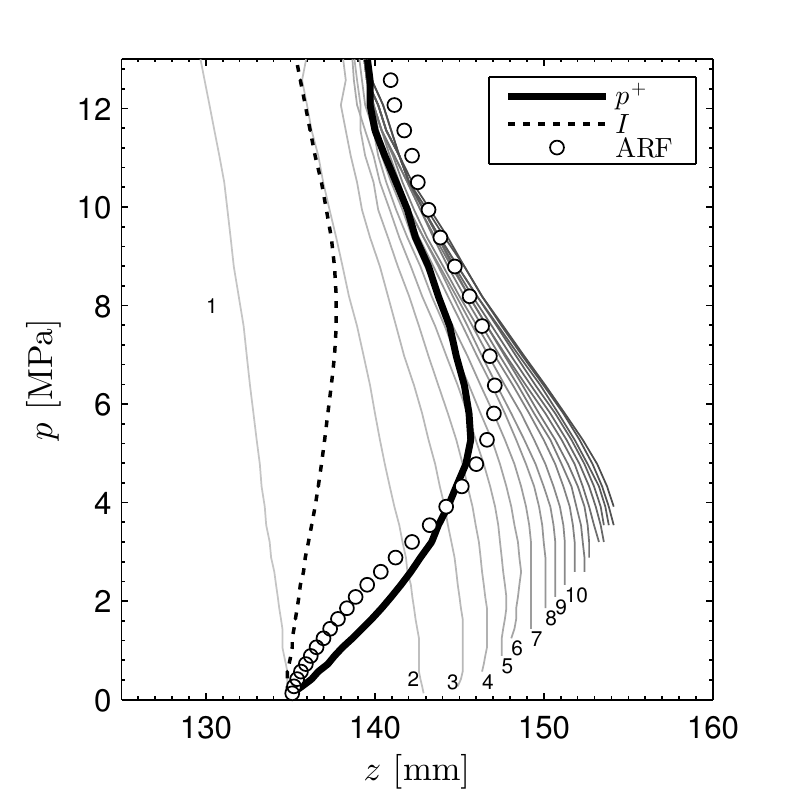}
	\caption{On axis maximum pressure, intensity and ARF location obtained from the KZK simulation of the ultrasonic beam. Grey lines, numbered, represent the on axis maximum pressure location of each harmonic.}
	\label{fig:5}
\end{figure}

In order to understand the on-axis behavior of the particular points of the beam, it has been performed a KZK simulation of the experiment in a range of excitation that covers from linear to strong nonlinear regime of propagation. Figure.~\ref{fig:5} shows the on-axis location of the peak value for the pressure ($p^+$), each harmonic (gray numbered lines), intensity and ARF. In linear regime, as intensity is proportional to the square of pressure and ARF is proportional to intensity the three maximum matches and all quantities peak at the same point. However, when moderately nonlinear regime is reached, harmonics begin to appear (gray lines in Fig.~\ref{fig:5}). Because the nonlinear generation is progressive and cumulative, and due to diffraction is less important at higher frequencies, harmonics focus farther away than the fundamental, and each one further than the previous one. Consequently, the peak of the total pressure moves away from the emitter. In this region, intensity also moves, but very slightly. The intensity of a nonlinear wave is the sum of the intensities of each harmonic, but as long as the main amplitude remains in the fundamental harmonic, it retains most of the energy and consequently sets the position of the maximum intensity. The relevance of the fundamental is enhanced in the calculation of the intensity because it is proportional to the square of the pressure. It does not happen the same with the ARF, which is also proportional to the intensity of each harmonic, but corrected by the absorption coefficient, as shown in Eq.(\ref{eq:ARFI}). In water, absorption increases with the square of the frequency, so the relevance of the harmonics is enhanced, moving significantly the peak of the ARF even in moderately nonlinear regime. As we increase the initial excitation amplitude of the emitter the number of harmonics in the propagation increase, which can cause even the ARF peak surpass the pressure peak. All these phenomena saturate (at around 6 MPa) and even retract at very high excitation levels because shock waves appear and nonlinear absorption reduces drastically the relevance of the harmonics. For this reason the peak locations of pressure, intensity and ARF move back to the emitter.

\subsection{Dynamic nonlinear focal shift in water}\label{Sec32}

\begin{figure}[tbp]
	\centering
	\includegraphics[width=8cm]{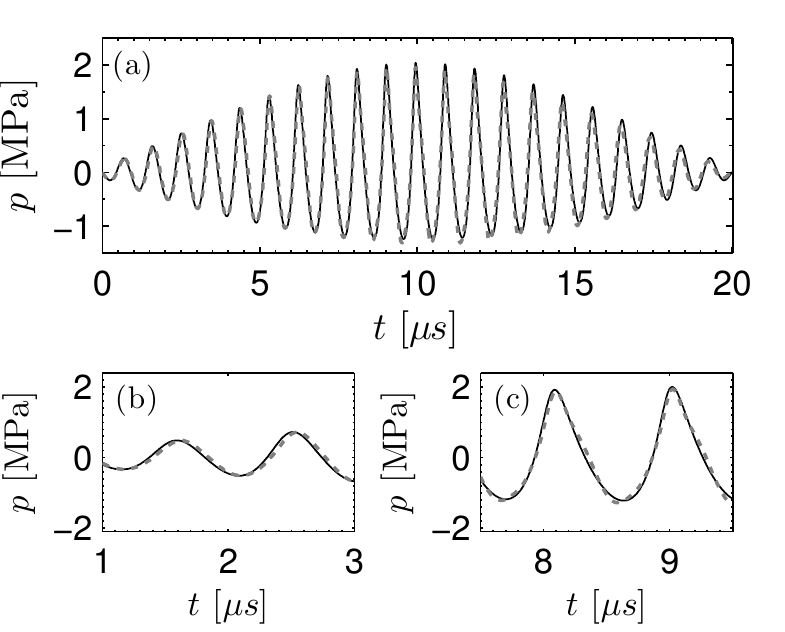}
	\caption{Experimental (dashed line) and KZK simulated (continuous line) waveform measured at the focus for the 25-kHz AM beam. The input voltage in experiment was 228 Vpp and the initial pressure in simulation was 105 kPa, chosen to produce equal pressure at the focus.}
	\label{fig:6}
\end{figure}

\begin{figure}[tbp]
	\centering
	\includegraphics[width=8cm]{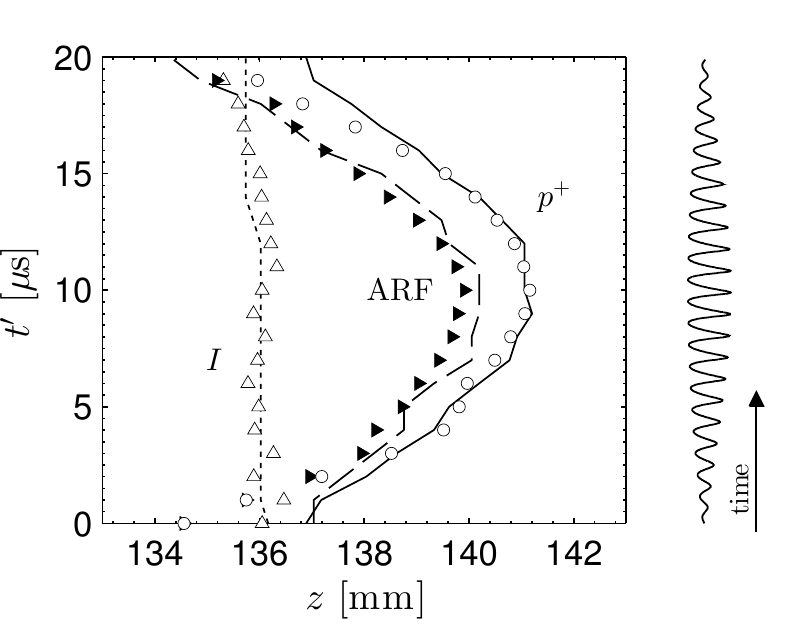}
	\caption{On-axis simulated (lines) and experimental (markers) of the peak pressure (white circles), maximum intensity (white triangles) and maximum ARF (black triangles) locations obtained for each period of the AM-carrier component. $t'$ is the dimensionless retarded time, where the time delay due to wave propagation distance has been eliminated.}
	\label{fig:7}
\end{figure}

The amplitude variation along the temporal profile of the 25 kHz-AM signal (see Fig.~\ref{fig:6}(a)) involves that the linear regime propagation for periods with small amplitude (without distortion; see Fig.~\ref{fig:6}(b)), and the nonlinear regime propagation for the central periods (with large amplitude and distortion; see Fig.~\ref{fig:6}(c)), coexist during the propagation of the wave. This results in dynamic focal shifts during the application of the signal, as it can be seen in Fig.~\ref{fig:7} for the peak pressure, intensity and ARF.

These results show that the focusing characteristics of the beam can change in time when we modulate the excitation signal. Figure~\ref{fig:7} represents the highest excitation value we have reached in the experiment: 228 $\mathrm{V}_p$ in the transducer and a peak pressure of 2.02 MPa at the focus. The dynamic process proceeds as follows: first, the low amplitude cycles of the AM beam focus on a point around 136 mm from the source according to the medium properties and the source physical characteristics: frequency, aperture, and geometrical focal length \cite{Makov2008}. Then, as the local amplitude of the AM signal reaches nonlinear regime in the central part of the AM packet (see Fig.~\ref{fig:6}(c)), the focal maxima moves away from the source. At this point, the displacement reach a maximum that perfectly agrees the displacements shown previously in Fig.~\ref{fig:3} and Fig.~\ref{fig:4}, and then returns to the transducer when the amplitude of the AM signal decreases. As shown, different behavior of the shift has been obtained for pressure, ARF and intensity. The maximum focal displacement observed was $+6.4$ mm for positive pressure, $-4$ mm for rarefaction pressure, $+1$ mm for intensity and $+4.6$ mm for the ARF. These results agrees to the nonlinear focal shifts obtained in Sec.~\ref{Sec32} for continuous beams. %, showing that when calculating the ARF, the term $c_0^{-2}\partial I/\partial t$ can be neglected for a $25$ kHz-AM signal \cite{guzina2015}.

\begin{figure}[tbp]
	\centering
	\includegraphics[width=8cm]{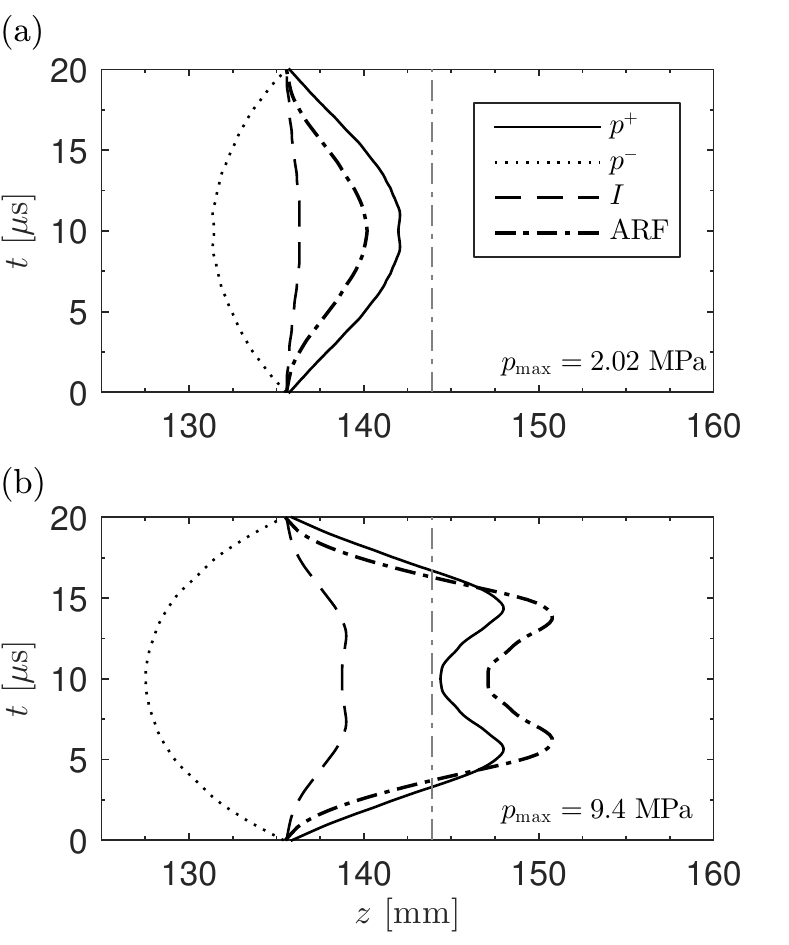}
	\caption{On-axis dynamic focal displacements in water for a 25 kHz AM signal. Up: moderate nonlinear regime. Down: strong nonlinear regime. Peak pressure (solid line), rarefaction minimum (dashed line), intensity (dotted line) and ARF (dashed-dotted line). The vertical dashed line shows the geometric focus. (a) Weakly nonlinearity scenario, (b) strongly nonlinear scenario.}
	\label{fig:8}
\end{figure}

As no shock waves are present in our experiment, we explore the strong nonlinear regime of propagation in an AM-modulated beam by means of a KZK simulation. Fig.~\ref{fig:8} shows the nonlinear shift in two scenarios (moderate nonlinear regime: 2.02 MPa at the focus, which correspond with the maximum excitation in our experiment) and 9.4 MPa at the focus (strong nonlinear regime).

The strong nonlinear scenario presents more complex dynamics than the moderate one. The maximum displacement of the focus was $+12.5$ mm for positive pressure, $-8.1$ mm for rarefaction pressure, $+3.4$ mm for intensity and $+15.1$ mm for ARF. The maximum position of the pressure exceeds the geometrical focus in $+4.2$ mm, and the ARF one in $+6.7$ mm. However, when the signal amplitude is high enough to produce shock waves in the propagation, the focus returns to the transducer in both cases. It happens when the central part of the AM packet is propagating through the media, and it is due to the effect of the nonlinear absorption, associated to shock waves, which saturate the harmonic generation processes. 

\subsection{Dynamic nonlinear focal shift in liver}\label{Sec33}

\begin{figure}[tbp]
	\centering
	\includegraphics[width=8cm]{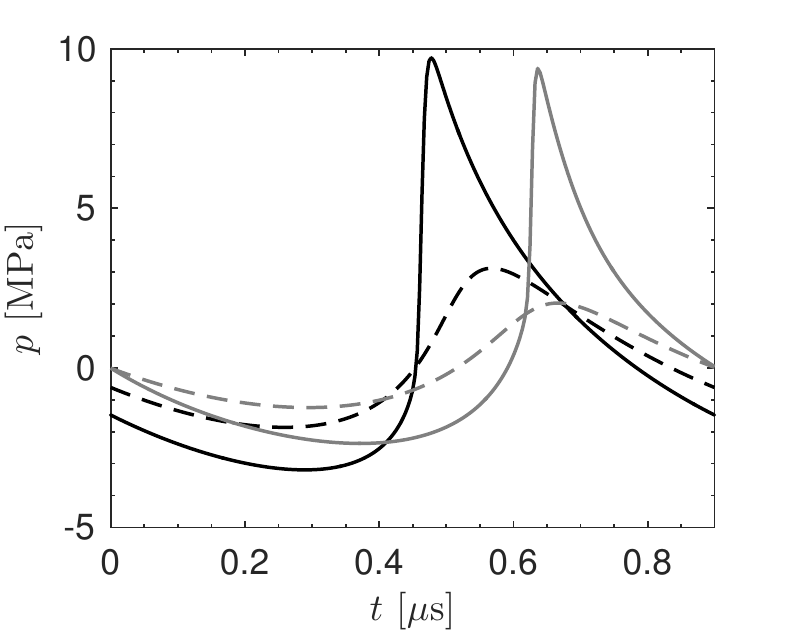}
	\caption{Comparison of the waveforms captured at the peak pressure location in water (gray lines) and human liver tissue (black lines). Two excitation levels per media are shown, a moderately nonlinear scenario (2.9 MPa), and a strongly nonlinear scenario (9.5 MPa).}
	\label{fig:10}
\end{figure}

Multiple relaxation and inhomogeneities were introduced in the KZK model to simulate the propagation of the beam through a layer of soft-tissue (human liver). A set of seventy simulations were performed in a range of excitation that covers from linear to the strongly nonlinear regime. Figure \ref{fig:10} shows the location of the peak pressure, each harmonic, intensity and ARF evaluated from Eq.(\ref{eq:ARFI}). First, it can be observed that in the linear regime the focus is shifted towards the source due to the strong attenuation, where all the magnitudes peak at $z=127.7$ mm (compared to 135.6 mm obtained in water). The linear position of the focus is located at 127.7 mm from the transducer, $-8.3$ mm before the focus obtained in water (136 mm). This effect is produced by the medium absorption in conjunction with an elongated focal area. Due to the high $f$-number of the source (2.57), the focal area, defined as the acoustic field area that overcomes $1/2$ of the peak intensity, has an axial length of $45\lambda$. During propagation through the focal area, where locally plane waves can be considered, the wave suffers from exponential decay. Over this path a total of 3.2 dB of attenuation is observed in tissue (considering 0.5 dB/cm/MHz the absorption of the liver\cite{Fujii2002}), so the peak pressure locates at the initial part of the focal area (127.7 mm) instead of at the point where it does in weakly absorbing media as water (136 mm). Despite this effect, in weakly nonlinear regime ($p^+<2$ MPa) the nonlinear behavior of the peak pressure and intensity in soft-tissue are almost similar than in thermo-viscous media, where each harmonic is progressively generated along the propagation and each one focus farther away than the previous one. However, the ARF shows a different behavior. ARF is also calculated by summing the intensities of the harmonics corrected each one by a factor proportional to its absorption. In soft-tissues the absorption follows a frequency power-law with an exponent $\gamma\approx 1$, (in contrast with thermo-viscous fluids where $\gamma=2$). Therefore, concerning the ARF, higher harmonics present less relative importance to the fundamental in tissue that in an equivalent thermoviscous media with $\gamma=2$. Due to that, the observed focal shift of the ARF in human-liver is relaxed compared to water, as shown in Fig.(\ref{fig:10}). In the present case the ARF peak location falls between the maximum intensity and peak pressure location. 

Then, in the moderately nonlinear regime ($2<p^+<4$ MPa), it can be observed that the beam self-refraction grows with amplitude up to $p^+<6$ MPa, a total shift of $+2.5$ mm was observed for the intensity, while it was $+9.4$ mm for the peak pressure and $+7.6$ mm for the ARF. Note these values are almost double of the corresponding ones in water.

\begin{figure}[tbp]
	\centering
	\includegraphics[width=8cm]{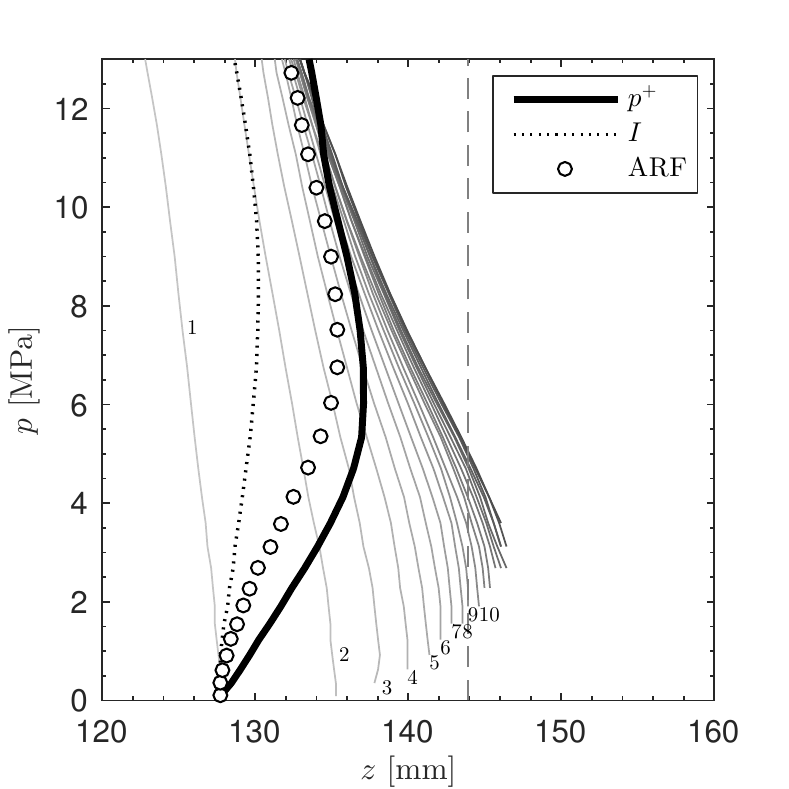}
	\caption{On axis maximum pressure, intensity and ARF location obtained from the KZK simulation of the ultrasonic beam in tissue. Grey lines, numbered, represent the on axis maximum pressure location of each harmonic.}
	\label{fig:9}
\end{figure}

As previously, 25 kHz-AM excitation signals were applied in two scenarios: moderate and strongly nonlinear propagation. The initial pressure values were chosen to produce similar pressure amplitudes at the focus as in the case of water: $p_0=0.26$ MPa amplitude excitation at the source surface, which leads to a peak pressure of $p^+2.9$ MPa (moderate nonlinear scenario); and $p_0=0.69$ MPa at the transducer surface, which leads to a peak pressure of $p^+9.5$ MPa (strongly nonlinear scenario). Waveforms captured at peak pressure location are shown in Fig.~\ref{fig:9} for both water and tissue simulations. Shock waves are formed for the highest amplitude calculations.

\begin{figure}[tbp]
	\centering
	\includegraphics[width=8cm]{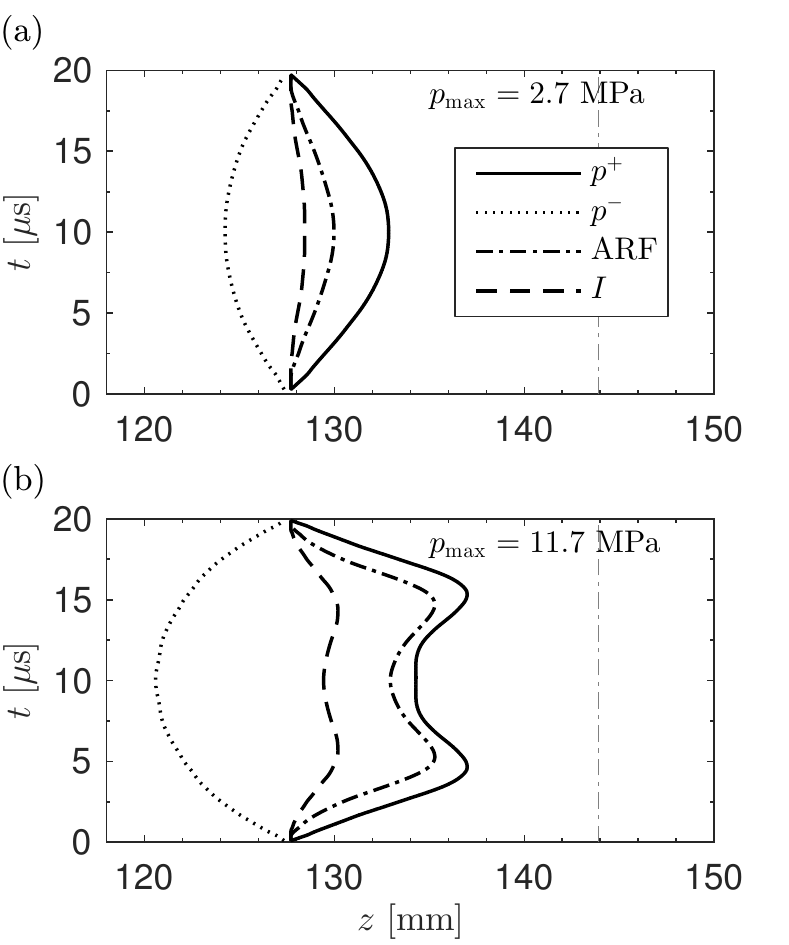}
	\caption{On-axis focal displacements for a 25 kHz-AM beam in liver. Peak pressure (solid line), rarefaction minimum (dashed line), intensity (dotted line). The vertical dashed line shows the geometrical focus. (a) Weakly nonlinearity scenario, (b) strongly nonlinear scenario.}
	\label{fig:11}
\end{figure}

Figure~\ref{fig:11} shows the dynamic location of the on-axis positive pressure, minimum rarefaction and intensity maxima. In the moderate scenario, as shown in Fig.~\ref{fig:11}~(a), the shifts are shorter compared to those obtained in water: $+5.4$ mm for peak pressure, $-3.5$ mm for minimum pressure, $+0.7$ mm for intensity and $+2.3$ mm for the ARF, i.e., 15\% lower for pressure, 30\% for the intensity, and 50\% for the ARF observed in water. The high value of the absorption in liver (0.5 dB/cm) compared to the absorption in water (0.002 dB/cm) reduces drastically the amplitude of the harmonics, and, consequently, the nonlinear shift effect.

Figure~\ref{fig:11}~(b) shows the dynamic nonlinear focal shift results for the strongly nonlinear scenario in human liver. A displacement of $+9.3$ mm for peak pressure is obtained, $-7.1$ mm for rarefaction pressure, $+2.5$ mm for intensity and $+7.6$ mm for the ARF. The maximum on axis pressure location does not exceed the geometrical focus because the linear focus is closer to the emitter (127.7 mm) than in water, as we explained above. In this case, the recoil in the shift is also observed in the central part of the AM packet, but smaller than in water: the strong absorption in tissue makes it more difficult to generate shock waves in the pre-focal region.

\section{Conclusions}\label{Sec4}
We have shown an analysis of the nonlinear behavior of the on-axis maximum pressure, intensity and ARF in a continuous and a 25 kHz AM beam (emitter characteristics: $f$-number = 2.6 and $f_0=1.112$ MHz). The pressure fields have been measured in water and compared to the ones obtained from numerical results based on the KZK equation, obtaining a good agreement between experimental and numerical data. 

As overall results, for a 25 kHz AM frequency, the maximum nonlinear shifts obtained in the AM beam match with the ones obtained in the continuous beam both, in linear and nonlinear regime; the on axis peak pressure moves away from the transducer as the excitation in the emitter increases, meanwhile the rarefaction pressure moves in the opposite direction; intensity suffers a slight displacement and ARF moves away in an interval between the intensity shift and the positive pressure shift in the moderate nonlinear regime, and surpass the pressure shift in the strong nonlinear regime. Once the shock waves begin to appear, the nonlinear shifts begin to decrease. These entire phenomena have been explained in terms of harmonic generation and absorption during the propagation in a lossy nonlinear medium both, for a continuous and an AM beam. 

The results presented demonstrate the ability of nonlinear AM beams generated with a mono-element transducer to produce dynamical axial focal displacements. Linear and nonlinear propagation regime, as well as the presence of shock waves and nonlinear absorption coexists during the semi-period of modulation of the beam (20$\mu$s in our case). The same results, but with reduced magnitude, have been observed in a KZK simulation of the beam propagating in human liver. 

Acoustic radiation force, and especially dynamic ARF, is widely used in many elasticity imaging techniques to induce displacements of tissue\cite{Maleke2006}, or to remotely induce shear waves inside the medium\cite{Nightingale2001}. We have observed that for amplitude modulated beams the position of ARF maximum changes with time following the modulation function. Although the focusing degree of the devices used in these imaging techniques is considerably higher than the one considered in our experiment, and as a consequence the nonlinear shift effects are minimized in these image techniques, they can be important enough to affect the calculation of the elastic characteristic of the tissue. A future research is going to be carried out in this line. Finally, the phenomenology presented in this work lead us to think in the possibility to explore the generation of supersonic shear waves by means of a mono-element transducer excited with the appropriate amplitude modulated signal: the frequency modulation would define the speed of the ARF displacement, meanwhile the maximum amplitude would fix the total distance covered.

\section*{Acknowledges}
FC acknowledge financial support from the program AICO/2016-108 of the Generalitat Valenciana. NJ acknowledge financial support from FPI grant PAID-2011 of the Universitat Polit\`ecnica de Val\`encia. NGS acknowledge financial support from projects DPI2013-42236-R of the Spanish Ministry of Economy and Competitivity, S2013/MIT-3024 of the Community of Madrid and from FPU grant 12/02187 of the Spanish Ministry of Education, Culture and Sport.

\bibliography{dynamicNonlinearFocalShift}

\end{document}